\documentclass[openacc]{rsproca_new}

\usepackage{tikz}
\usepackage{amsmath}
\usepackage{pgf} 
\usepackage{tabularx}   
\usetikzlibrary{arrows.meta, automata, positioning, quotes}
\usepackage{multirow}

\begin{document}

\title{Using symbolic networks to analyse dynamical properties of disease outbreaks}

\author{
Jos\'e L. Herrera-Diestra$^{1,2}$, Javier M. Buld\'u$^{3,4,5}$, Mario Chavez$^{6}$ and Johann H. Mart\'inez$^{8,1,3}$}

\address{
$^{1}$ICTP South American Institute for Fundamental Research, Instituto de F\'{\i}sica Te\'orica, Universidade Estadual Paulista, S\~ao Paulo, Brazil\\
$^{2}$CeSiMo, Facultad de Ingenier\'ia, Universidad de Los Andes, Venezuela\\
$^{3}$Grupo Interdisciplinar de Sistemas Complejos (GISC), Madrid, Spain\\
$^{4}$Laboratory of Biological Networks, Center for Biomedical Technology - UPM. Madrid, Spain\\
$^{5}$Complex Systems Group, Universidad Rey Juan Carlos, M\'ostoles, Spain\\
$^{6}$CNRS UMR7225, H\^opital Piti\'e Salp\^etri\`ere, Paris, France\\
$^{7}$INSERM-UM1127, Institute du Cerveau et de la Mo\"elle Epini\`ere. H. Salp\^etri\`ere, Paris, France\\
$^{8}$Universidad de los Andes, Biomedical Engineering Department, Bogot\'a, Colombia
}

\subject{Complex Systems, Information Theory, Time Series, Symbolic dynamics}


\corres{Jos\'e L. Herrera-Diestra\\
\email{diestra@gmail.com}}

\begin{abstract}
We introduce a new methodology to analyze the evolution of epidemic time series, which is based on the construction of epidemic networks. First, we translate the time series into ordinal patterns containing information about local fluctuations of the disease prevalence. Each pattern is associated to a node of a network, whose (directed) connections arise from consecutive appearances in the series. The analysis of the network structure and the role of each pattern, allows classifying them according to the enhancement of entropy/complexity along the series, giving a different point of view about the evolution of a given disease.
\end{abstract}

%

\maketitle
\section{Introduction}
Time Series Analysis (TSA) is a successful field of a cross disciplinary character. In economics, climate, geophysics and 
statistics \cite{GRANGER1986,liming2013, NOAKES1988,KIM2016}, among others, 
forecasting based
on the observation of the previous states of the system is of critical interest. Engineering and communications 
are focused on control and parameter detection of signals\cite{WERON2014,NOURY2019,Rowsseeuw2019}. 
Classification, clustering or detection of abnormalities in a collection of samples, are the main targets for Data mining and Machine learning\cite{SI2018,HUSSAIN2018}. A 
time series is basically the collection of a given variable describing the activity of a system during different time-points. This system may be
linear or nonlinear, translating this properties to the observed time series. 
Regardless of the diversity of techniques to acquire and study a collection of temporal samples, the extraction of information about the underlying dynamical system from the 
observation of the evolution of one (or several) of its variables, is frequently a non-trivial task \cite{Fu2011,DeGooijer2006}.

Notwithstanding the difficulties, several methodologies and tools have been proposed to better capture and understand the properties of a system from the observation of its dynamics. 
In general, they rely on the fact that time series have internal structures 
along several dimensions (time, amplitude, phase, frequency, ...) that contain information about the dynamical response of a system. The way these structures are related 
would help at describing the system's activity and predicting its evolution. With this regard, TSA has broadly focused on two perspectives: One in which few parameters can 
describe time series if they come from stationary stochastic processes; and a non-parametric one, in which the estimation of the spectral density or higher order conditional 
moments totally describes a collection of samples\cite{CHEN1997,Kocsis2017}. In the other perspective, TSA methods can be grouped in univariate and multivariate ones. The latter 
accounts for techniques to quantify the contribution of two or more variables on a single event, while the former aims at the describing and inferring the 
evolution of temporal structures based on a single variable\cite{HE2015,HOGA2017,ABOAGYESARFO2015,SALLES2019274}. Regarding univariate methods, a central 
question is to know whether temporal structures correlate each other. The serial correlation quantifies the point-by-point
correlation of a signal with a delayed version of itself\cite{gubner2006}. Autocorrelation is then useful to accurately seek for repetitive patterns in a signal. 
However, when a signal is split into consecutive segments, autocorrelation fails in capturing the interplay between them. Therefore, a naive but not trivial question may raise about whether 
signal segments of a time series might be related to each other.

In this context, this paper is focused on the hypothesis that consecutive segments of a time series may be inter-relayed, transferring information along them. To test this hypothesis, we unravel 
how segments transmit information between them, with the final objective of gaining additional insights about what are the properties of time series and, ultimately, of the system behind them. 
We quantify the communication levels among consecutive segments of a time series using a combination of 
Networks Science (NS) and Symbolic Dynamics (SD). 

The use of NS has reached a broad range of areas, such as social sciences, biology, ecology, neuroscience, epidemics, among others \cite{newman2010}. 
Its main advantage relies on the ability of translating almost any system under interaction into nodes, which are endowed with properties that can be extracted from almost any variable of the system. 
These nodes are connected by links, entities that account for any kind of interaction between them.
The mathematical background of NS is useful to design artificial models leading to different network structures, while its transversal nature allows these models to be 
applied to different kind of datasets \cite{newman2010,estrada2015}. NS has also been applied in data mining and machine learning 
problems, and more recently it has been implemented as a concomitant tool of TSA \cite{ZANIN2016,CAMACHO2018,TANIZAWA2018,WANG2019,ZOU2019,Lacasa2008,Iacovacci2019,Masoller2015}.

In this context, publications using {\it visibility graphs} have introduced new perspectives on how to extract information from experimental time series \cite{LozanoPerez79,Lacasa2008}. 
The idea behind visibility graphs is that a time series can be transformed into a network, where nodes are the different values of the time series and their links are created between any two amplitudes as long as they can ``see" each other without being covered by another intermediate sample, as if it were a series of concatenated hills and valleys \cite{Lacasa2008}.  
Since the seminal work of Lozano-Perez et. al.\cite{LozanoPerez79}, visibility graphs have been applied to a broad 
range of problems, from planning collision-free paths to avoid polyhedral obstacles, characterization of random time series, 
combinatorics on words, to three dimensional perspectives for image processing \cite{Nunez2012}.

Alternative techniques to map time series dynamics into a graph representation have been introduced by 
means of symbolic dynamics. SD assumes a symbol as a perceptible idea that encloses common characteristics of a set of 
samples in a time series. In other words, a symbol contains valuable information about a temporal segment but, at the same time, 
it simplifies the analysis of the original variable. The symbol definition has lead to 
different approaches, many of them with interesting results \cite{Lin2003,Kennel2004,Daw2000,Martinez2018}. However, 
among the diversity of definitions of a symbol, the one proposed by Bandt and Pompe is nowadays the most endorsed by
the community of researchers making use of symbolic time series \cite{Bandt2002}. The originality of this approach consists of a formalism that quantifies the information of a 
time series by its transformation into ordinal patterns, which take into account the ranking between consecutive values.

An ordinal pattern $\pi$ is defined by comparing the relative amplitudes of a temporal segment of $D$ consecutive elements observations ($D$ is also called the dimension of the pattern), and mapping
 them into a one-dimensional ordinal space $\{0,1,...,(D-1)\}$, where the highest value in the segment will be tagged by $D-1$ and the 
lowest will be assigned $0$, with the rest of the values going in descending other. Thus, each element of the pattern $\pi$ only contains the order, disregarding its specific value. 
For finite time series of length $M$, when the 
 cardinality $D$ of the temporal segments is fixed, the amount of possible symbols is determined by $D!$. This way, 
 the variables reduction boosts the run-time of exploratory experiments. Next, the probability distribution of
  finding a pattern $\pi$ in the symbol sequence is obtained and analysed in order to obtain information about the general properties of the underlying system. This methodology has been proved to be robust against noise, fully data-driven, adequate 
  under weak stationarity and computationally efficient with no further assumptions over the 
  datasets \cite{Martinez2018,Monetti2013,amigo2007,amigo2015,zanin2008,cazelles2004,rosso2007}. It
   has been extensively used to characterize time series of different nature, as well as for mapping a collection 
   of samples into a directed graph. 
   With this regard, a directed graph containing $D!$ nodes can be created when different symbols $\pi$ of a 
   signal are consecutively connected as they appear in the time series \cite{Masoller2015}. Using this transformation of patterns into networks, 
   lasers and chaotic time series were characterized by means of the link entropy and 
    a non-normalized version of Shannon entropy of the resulting nodes. This work revealed the usefulness of mapping time series into a graph, and 
    opened the door to further studies about the structure of the resulting networks.

In this paper, we go one step beyond by studying the transfer of information along temporal segments of epidemic time series. We transform them into ordinal patterns, which are then
projected into nodes of a network whose links are based on consecutive appearances along the time series.
In particular, we construct symbolic networks from epidemic time series ($x_t$) of vector-borne (Dengue, Malaria) and air-borne (Influenza) 
diseases reported at different countries (see Tab. \ref{tab:01} of Methods Section). We also introduce a family of five novel parameters that characterize the role of the nodes (patterns) of 
the network, specifically: 1) the amount of entropy entering/leaving a node $\pi$, by means of the \textit{incoming/outgoing entropy} ($H_{in}, H_{out}$), 
2) the amount of complexity entering and leaving a node $\pi$, by means of the \textit{incoming/outgoing complexity} ($C_{in}, C_{out}$),
3) the level of conductance of information of a node, by 
 means of the \textit{flux} of entropy/complexity ($\phi_{H,C}$), 
 4) how much a node amplifies or attenuates  entropy/complexity ($A_{H,C}$), and 5) the internal \textit{fluctuation} ($f$) inside each pattern.
Using these five metrics we are able to identify differences between  diseases and assign specific roles to the different patterns of the time series. 
We explore the interplay between the fluctuation $f$ of each pattern $\pi$ and the role the corresponding node has on the network structure.
Next, we compare our results with a set of synthetic outputs of different complexity in order to infer the closeness similarities of 
these diseases with synthetic models. Our results demonstrate that this methodology unveils differences among air-borne and vector-borne diseases, and evidences how 
these outbreaks share similarities with autoregressive processes of linear and nonlinear nature.

\section{Methods}
\subsection{Datasets}
Epidemiological cohorts are composed of nine time series $x_t$ of different length $M$ corresponding to three diseases in six different countries for the time periods shown in Tab. \ref{tab:01}. Each times series represents the weekly amount of individuals infected with a specific disease. In order to account the non-stationarity of signals, we extracted their log-returns $y_t=log(x_t-1)-log(x_t)$.
\begin{table}[!h]
\centering
\caption{Datasets: Countries and diseases. For each country and disease, we show the number of weeks $M$ of each available epidemiological time series with the corresponding time period (in years) in parenthesis.}
\label{tab:01}
\begin{tabular}{l c c c}
 \hline
 {\bf Country}  &  {\bf Dengue}      & {\bf Influenza}    & {\bf Malaria}      \\ \hline
 {\it Australia}&                    &  974 $(1997-2015)$ &                    \\ 
 {\it Colombia} & 626 $(2005-2016) $ &                    &626 $(2005-2016)$  \\ 
 {\it Japan}    &                    & 964 $(1998-2015)$ &                    \\ 
 {\it Mexico}   & 678 $(2000-2015)$ &  830 $(2000-2015)$ &                    \\ 
 {\it Singapore}& 838 $(2000-2015)$ &                    &                    \\ 
 {\it Venezuela}& 660 $(2002-2014)$ &                    & 669 $(2002-2014)$ \\
 \hline
\end{tabular}

\end{table}

The datasets analyzed in this paper were extracted from online reports of corresponding {\it Ministry of Health} of each country. \cite{australiaER,colombiaER,japanER,mexicoER,singaporeER,venezuelaER}. 
We studied diseases in terms of the graph representation of their aggregated time series. Time series are cut into vectors of specific length $D$, where each different vector corresponds to a node of the network. Nodes (vectors) are then consecutively connected in order of appearance, which leads to a directed graph. Patterns associated to each node are extracted from a symbolic transformation mapping the inner relative amplitudes of each vector into an ordinal representation. When two vectors co-occur sequentially more than one time, a weight is given to the direct connection between the two vectors (nodes), leading to a weighted graph. Next, we analyze networks of diseases using a family of novel network parameters, consisting on a set of information-based metrics quantifying the entropy and complexity (\textit{in/out entropy}, \textit{in/out complexity}) of the nodes and a set of parameters assessing their dynamical role (\textit{flux}, \textit{amplitude} and \textit{fluctuation}).

\subsection{Symbolic transformation}
\begin{figure}[ht]
\centering
\includegraphics[width=.8\linewidth]{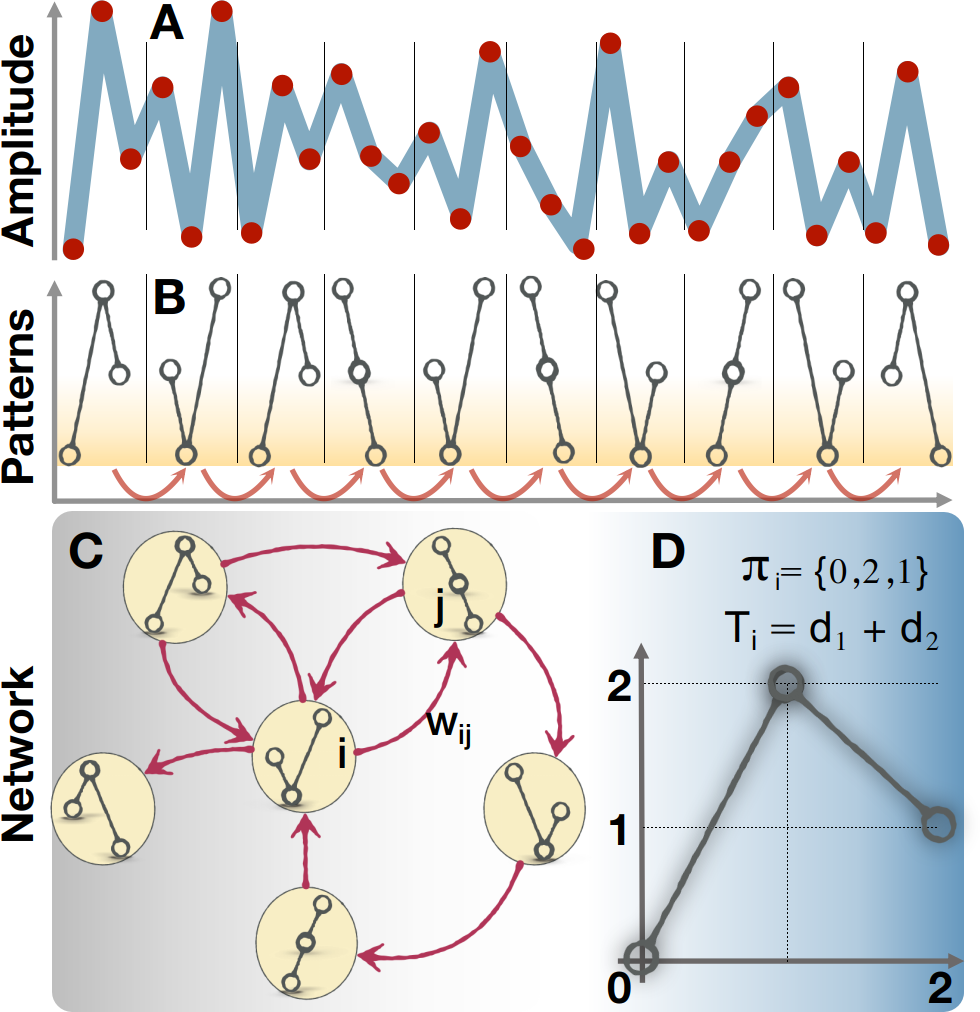}
\caption{Extracting ordinal patterns from a time series. {\bf A.} Signal amplitudes are splitted into consecutive vectors of length $D$ (in this example, $D=3$). {\bf B.} The values of the amplitude inside each vector are translated into a ranking. Consecutive patterns are connected through a direct link. {\bf C.} Directed graph derived from the time series and patterns. {\bf D.} For illustrative purposes, vector $\{-0.3,1.2,0.5\}$ is transformed into the ordinal pattern $\pi=\{0,2,1\}$ and its variability is obtained as $T_i=d_1+d_2=\sqrt{5}+\sqrt{2}=3.65$ (where $d_1$ and $d_2$ are the lengths of the lines connecting two consecutive rankings), corresponding to a fluctuation of $f=2$ (see Tab. \ref{tab:02} and main text, for details).}
\label{fig:01}
\end{figure}

Following the methodology proposed by Bandt \& Pompe \cite{Bandt2002}, we retrieve all different patterns $\pi$ of length $D$ appearing in a signal $x_t$. When defining an embedding dimension $D$, we are constrained by the size $D$ of the patterns, since $D!$ is the total number of possible patterns, e.g., $D=3$ leads to $D!=6$ possible symbols emerging from $x_t$. The amount of all appearing symbols is also closely related to the length $M$ of the time series, since extremely short time series have not enough statistics to guarantee a fully resolved distribution of patterns. For a statistical reliable estimation, we follow the condition $M-D \gg D!$ \cite{Tiana2010}. Once this condition is fulfilled, each symbol is then constructed by considering consecutive samples of length $D$  extracted from $x_t$. Under this mapping, $x_t$ ($\forall$ $t=1,2,...M$), is transformed into a restricted number of patterns encoding the relative inner amplitudes of the $D-$dimensional vector $\{x_t, x_{t+1}, ..., x_{t+D}\}$. Samples are arranged (or ranked) onto the permutation $\pi=(\pi_0, \pi_1, ..., \pi_{(D-1)})$ of $(0,1, ..., D-1)$ fulfilling $x_{t+\pi_0}\leq x_{t+\pi_1}\leq x_{t+\pi_{D-1}}$. Hence, each permutation is a symbol of the full spectrum of available patterns.  The full process is summarized, schematically, in  Fig. \ref{fig:01}. This methodology had been proven to be computationally efficient, fully data-driven with no further assumptions over the data, robust against noise, and well-behaved under weak stationarity \cite{amigo2007,amigo2015,zanin2008,amigo2010,keller2014,cazelles2004,johann2018,zanin2012,rosso2007}.

\paragraph{Pattern Fluctuation.}
We propose to quantify the internal variations of each symbol by defining its \textit{fluctuation} (\textit{f}). With this aim, we quantify the total length $T$ of the lines connecting the ranking values inside each pattern, as shown in Fig. \ref{fig:01}D. Note that, the larger the value of $T$, the higher the variability inside each pattern. Next, define the fluctuation $f$ of a given pattern as the ranking of its corresponding value of $T$, from the lowest ($f=1$) to the highest (See Tab. \ref{tab:02} for details). In this way, we can group patterns in terms of their internal fluctuations, which gives interesting information about the temporal dynamics of the pattern. For instance, monotonic and periodic time series results in an abundance of fluctuations with low $f$, a parameter that can be included to complement the information about the specific features of a given node. 

\begin{table}[!h]
\centering
\caption{Pattern fluctuations for $D=3$. First row: All possible patterns $\pi$ of length $3$. Second row: The internal variability $T$ (obtained as the length of the connecting lines). Third row: The corresponding fluctuation parameter $f$. Note that different patterns have the same fluctuation, since they have the same internal varaibility.}
\label{tab:02}
\begin{tabular}{lcccccc}
\hline
{\bf $\mathbf{\pi}$} 	 

& \begin{tikzpicture}
\draw[thick,-] (0,0) -- (.5,.5);
\draw[thick,-] (.5,.5) -- (1,1);
 \filldraw[fill=gray] (0,0) circle[radius=2pt];
 \filldraw[fill=gray] (.5,.5) circle[radius=2pt];
 \filldraw[fill=gray] (1,1) circle[radius=2pt];
\end{tikzpicture} 

& \begin{tikzpicture}
\draw[thick,-] (0,1) -- (.5,.5);
\draw[thick,-] (.5,.5) -- (1,0);
 \filldraw[fill=gray] (0,1) circle[radius=2pt];
 \filldraw[fill=gray] (.5,.5) circle[radius=2pt];
 \filldraw[fill=gray] (1,0) circle[radius=2pt];
\end{tikzpicture}        

& \begin{tikzpicture}
\draw[thick,-] (0,1) -- (.5,0);
\draw[thick,-] (.5,0) -- (1,.5);
 \filldraw[fill=gray] (0,1) circle[radius=2pt];
 \filldraw[fill=gray] (.5,0) circle[radius=2pt];
 \filldraw[fill=gray] (1,.5) circle[radius=2pt];
\end{tikzpicture}

& \begin{tikzpicture}
\draw[thick,-] (0,.5) -- (.5,0);
\draw[thick,-] (.5,0) -- (1,1);
 \filldraw[fill=gray] (0,.5) circle[radius=2pt];
 \filldraw[fill=gray] (.5,0) circle[radius=2pt];
 \filldraw[fill=gray] (1,1) circle[radius=2pt];
\end{tikzpicture}

& \begin{tikzpicture}
\draw[thick,-] (0,.5) -- (.5,1);
\draw[thick,-] (.5,1) -- (1,0);
 \filldraw[fill=gray] (0,.5) circle[radius=2pt];
 \filldraw[fill=gray] (.5,1) circle[radius=2pt];
 \filldraw[fill=gray] (1,0) circle[radius=2pt];
\end{tikzpicture}

& \begin{tikzpicture}
\draw[thick,-] (0,0) -- (.5,1);
\draw[thick,-] (.5,1) -- (1,.5);
 \filldraw[fill=gray] (0,0) circle[radius=2pt];
 \filldraw[fill=gray] (.5,1) circle[radius=2pt];
 \filldraw[fill=gray] (1,.5) circle[radius=2pt];
\end{tikzpicture} \\  \hline
{\bf $\mathbf{T}$}   & 2.828 & 2.828 & 3.650 & 3.650 & 3.650 & 3.650 \\  
{\bf $\mathbf{f}$}& 1 & 1 & 2 & 2 & 2 & 2  \\  \hline
\end{tabular}
\end{table}

\paragraph{Network construction.}
We built the network representation of the time series by transforming each different pattern into a node, following the methodology proposed 
by Masoller et al. \cite{Masoller2015}. 
Consecutive symbols were connected through a direct link (see red arrows of \ref{fig:01}B) and auto-loops (links leaving and reaching the same node) were dismissed. In this way, sequential associations of patterns lead to a directed graph, while repetitions of pattern co-ocurrence (e.g., when pattern $\pi_i$ and $\pi_j$ appear one after the other more than once) reinforce the weight $w_{ij}$ of the link going from node $i$ to $j$ (or, in other words, from pattern $\pi_i$ to $\pi_j$). The weight of a link $\pi_i\rightarrow \pi_j$, $w_{ij}$, is the relative number of times, in the sequence, the symbol $i$ is followed by symbol $j$; in this way, the link weights are normalized in each node, i.e. $\sum_{j=1}^{D!}w_{ij}=1$. Figure \ref{fig:01}C, shows the resulting network from the time series and patterns.

\begin{figure}[ht]
\centering
\includegraphics[width=.6\linewidth]{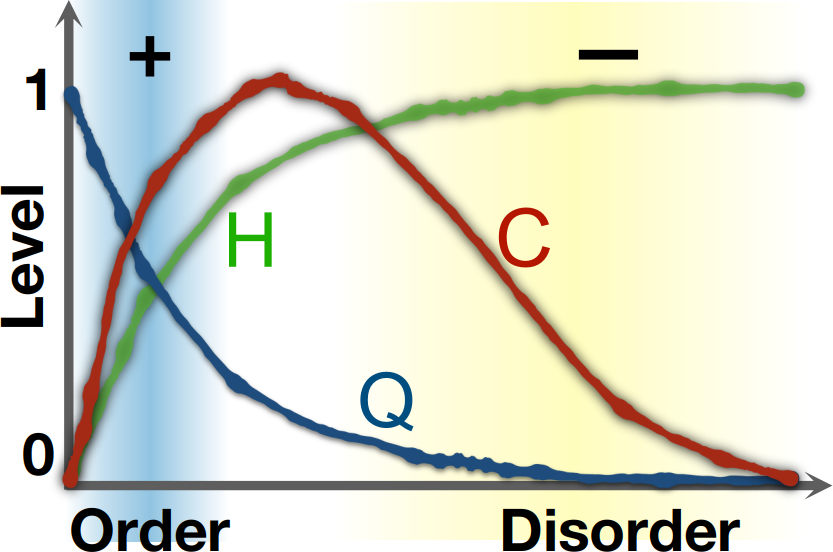}
\caption{
Qualitative representation of the entropy-complexity ($H\times C$) plane. Entropy $H$ increases with the disorder of the signal. Disequilibrium $Q$ decays as the signal approximates to a complete disordered state. Statistical complexity $C=Q\cdot H$ reaches its maximum when the system stays between order and disorder. Note the region of positive correlations between $H$ and $C$, which is highlighted in blue, while the region of negative correlation is highlighted in yellow.}
\label{fig:02}
\end{figure}

\subsection{Global measures}
Entropy $H$ and statistical complexity $C$ are useful when characterizing the appearance of ordinal pattern populations along time series \cite{rosso2007,zanin2012}. A simple diagram of these measures globally characterizes its dynamics (see Fig \ref{fig:02}B). In this way, a region of positive correlations between $H$ and $C$ reveals that the system is close to an order phase, meanwhile a negative correlation arises in a region of high disorder \cite{amigo2010} (see Fig \ref{fig:02}B). The [$H,C$] plane has been widely used for the description of a diversity of natural and artificial systems, always departing from the transformation of time series into ordinal patterns. \cite{amigo2007,amigo2010,amigo2015,zanin2008,keller2014,cazelles2004,carpi2010,zanin2012}

$H[p]$ and $C[p]$ are defined from the empirical distribution $p$ of ordinal patterns extracted from $x_t$. Permutation entropy $H[p]$  is given by the ratio between the Shannon entropy $S[p]$ and the maximum entropy $S_{max}=S[p_e]$, being $p=\{p_i\}$ the distributions of $i$ patterns and $p_e$ a uniform probability. In this way, $0\leq H[p]\leq 1$, with $H$ = 0 if $p_{i}=1$ and ${H}=1$ if ${{p}_{i}}=1/D!$ $\forall i$. The disequilibrium $Q$ evaluates the existence of preferred states among the accessible ones. It is defined as $Q[p]=Q_0D[p,p_e]$, with $Q_0$ being a normalization constant and $D[p,p_e]$ as the mean information for discriminating between $p$ and $p_e$ per observation $p_i$, defined as the symmetric form of the Kullback-Leibler relative entropy \cite{Kullback1951}. Low values of $Q[p]$ imply a distribution of patterns $p$ close to what one may expect at random, i.e. $p_e$. Otherwise, high values of $Q[p]$ reflect the existence of some privileged patterns. Finally, statistical complexity $C[p]=H[p]\cdot Q[p]$ quantifies the interplay between order and disorder of the system. Note that $C[p]$ vanishes if the system is close to the equilibrium (maximum disorder and minimum distance to uniform probability) or in a state of high order (minimum entropy). The triad of dynamical properties $\{H,Q,C\}$ is commonly known as the generalized statistical complexity measures of a time series. Here, we propose to use these global parameters capturing the properties of the time series of a system and combine them with the topological properties of directed-weighted graphs.

\subsection{Local measures}
Additionally to the pattern fluctuation $f$, we also propose a collection of node features based on global parameters. We define the incoming content of information of a node $i$ as the probability distribution $P^i_{in}=\{ p_l\}$ of the inbound links, $\forall$ $l=\{1, ..., k_{in}\}$, where $k_{in}$ is the in-degree. To do that, the weight of each incoming link is divided by the total weight of the $l$ links entering to $i$, such that $\sum_{l=1}^{k_{in}}p_l=1$. Reciprocally, we construct the outgoing information content $P^i_{out}=\{p_m\}$ upon outgoing $m$ edges from $i$ $\forall$ $m=\{1, ..., k_{out}\}$, with $k_{out}$ as the out-degree. All departing weights are divided by the total sum of its $m$ outbound weights fulfilling $\sum_{m=1}^{k_{in}}p_m=1$. This turns a node $i$ into an entity endowed with the property of conveying information among ordinal patterns, which can be interpreted as the way temporal structures (patterns) communicate (or transmit information) along time.

\begin{table}
\caption{Definitions of the node entropies based on the distribution of incoming and outgoing links of a node $i$.}
\label{tab:03}

\centering

\rule{\textwidth}{.1pt}
\vspace{-0.9cm}

\begin{subequations}

\begin{flalign}
& \text{Nodal entropy} & H_{in}[p_l] = \frac{S_{in}}{S^{max}_{in}}
&& H_{out}[p_m] = \frac{S_{out}}{S^{max}_{out}} && \label{equ:h}
\end{flalign}

\begin{flalign}
& \text{Nodal complexity} & C_{in}[p_l] = H_{in}[p_l] \cdot Q_{in}[p_l]
&& C_{out}[p_m] = H_{out}[p_m] \cdot Q_{out}[p_m] && \label{equ:c}
\end{flalign}

\vspace{-0.1cm}
\end{subequations}

\rule{\textwidth}{.1pt}

\end{table}

\paragraph{Nodal entropy and complexity.}
Under this framework, we can measure the Shannon entropy arriving to node $i$ as $S_{in} =  -\sum_{l=1}^{k_{in}} p_l\ln\left(p_l\right)$, and we can contrast it with its maximal entropy $S_{in}^{max}=log(k_{in})$ obtained for the uniform distribution $p_e=\{p_{l,e} \mid p_{l,e}=1/k_{in} ~\forall ~l\}$. The ratio between $S_i$ and $S_{in}^{max}$ leads to the \textit{incoming node entropy} $H_{in}[p_l] = S_{in}/S^{max}_{in}$ that is bounded by $0\leq H_{in}[p_l] \leq 1$. When only one link arrives to node $i$, i.e. $P^i_{in}=1$, the incoming node entropy is the lowest ($H_{in}=0$). On the other hand, when all possible links arrive to $i$ (i.e.,  node $i$ has $l=D!-1$ incoming links), the incoming node entropy is the highest  ($H_{out}=1$) if all incoming links have the same probability $p_l=1/k_{in}$. Analogous reasoning applied to the $m$ outgoing links leads to the {\it outgoing node entropy} $H_{out}[p_m]$. First row of Tab. \ref{tab:03} summarizes the definitions of the incoming and outgoing node entropies.

Similarly to global measures, comparing to a uniform distribution $p_e$ one can measure the disequilibrium of incoming links $Q_{in}[p_l]=Q_0\cdot D[p_l,p_e]$. Here, $Q_0=-2\{ \frac{k_{in}+1}{k_{in}}log(k_{in}+1) - 2log(2k_{in}) + log(k_{in}) \}^{-1}$ is the normalization constant leading to $0\leq Q_{in}[p_l] \leq 1$. $D[p_l,p_e]$ accounts for the Jensen-Shannon divergence defined in terms of $S_{in}$ as $D[p_l,p_e]=S_{in}[(p_l+p_e)/2]-S_{in}[p_l]/2-S[p_e]/2$. 

By multiplying $H_{in}$ and $Q_{in}$ we finally obtain the \textit{incoming node complexity} that is bounded between $0\leq C_{in}[p_l] \leq 1$. Following the same procedure as node entropy, we can define the {\it outgoing node complexity} $C_{out}[p_m]$ considering the $m$ outgoing links. Second row of Tab. \ref{tab:03} summarizes the definitions of the incoming and outgoing node complexities.

The previous parameters shed light on how temporal structures in time series are temporally correlated to each other, and how different patterns are related the increase of entropy and complexity along a time series.

\paragraph{Flux and Amplification of information.}
The flux is a vectorial quantity that describes the magnitude and direction of the flow of information passing through a node $i$. In directed graphs, direction is trivially associated with incoming links of node $i$ to its outgoing ones. Therefore, the net amount of ``information'' passing through $i$ can be quantified in terms of two nodal properties: 
The \textit{entropy $\phi_{H}$ and complexity $\phi_{C}$ fluxes}. Fluxes $\phi_{H}$ and $\phi_{C}$  are measures of the dynamical importance of a node, since they characterize which patterns behave as dynamical hubs by allowing that larger amounts of information entering and leaving the pattern $i$. They quantify the dynamical relevance of nodes by taking into account the entropies and complexities of the incoming and outgoing links. Table \ref{tab:04} contains the mathematical definitions of the entropy and complexity fluxes. 

\begin{table}[ht]
\caption{Definitions of the flux $\phi$ and amplification $A$ of a node $i$, both for the entropy and complexity of the incoming and outgoing links.}
\label{tab:04}

\centering
\rule{\textwidth}{1pt}
\vspace{-0.9cm}
\begin{subequations}

\begin{flalign}
& \text{Flux} & \phi_H = \sqrt{H_{in}^2 + H_{out}^2}
&& \phi_C = \sqrt{C_{in}^2 + C_{out}^2} && \label{equ:f}
\end{flalign}

\vspace{-\baselineskip}
\begin{flalign}
& \text{Amplification} & A_{H} = \frac{H_{out}^i}{H_{in}^i}
&& A_{C} = \frac{C_{out}^i}{C_{in}^i} && \label{equ:a}
\end{flalign}

\vspace{-0.1cm}
\end{subequations}

\rule{\textwidth}{1pt}

\end{table}

Importantly, fluxes allow to identify \textit{basin nodes}, which retain all incoming information from other patterns and do not transmit information to another ones, i.e. nodes whose entropy (or complexity) flux is ($\phi_H=\sqrt{H_{in}^2 + 0^2}=H_{in}$). On the other hand, \textit{generator nodes} have no incoming links, but convey information to another nodes ($\phi_H=\sqrt{0^2 + H_{out}^2}=H_{out}$). However, note that generator nodes may not exist in our kind of networks, since all patterns appear after a previous one (with the unique exception of the first pattern of the time series). Finally the \textit{flux dynamical hubs} are those  patterns receiving large amounts of information from other temporal structures (i.e., patterns that are reached from a large amount of different patterns), and redistribute it equally (i.e., also lead to a diversity of patterns). For example, in the extreme case of ($\phi_H=\sqrt{1^2 + 1^2}=\sqrt{2}$) we would have the most important hub for the entropy flux. As a consequence, by definition, the value of fluxes are bounded by $0< \phi_{H,C} \leq \sqrt{2}$.

\begin{table}[!h]
\centering
\caption{Definitions of the dynamical roles of a node. The first row depicts five types of transitions for a pattern $\pi_i$. $P_l$, $P_m$ are the probabilities associated to incoming and outgoing links. The second row corresponds to the role assigned to the nodes. The third row contain the limits and domains for the \textit{flux} $\phi_{H,C}$. Note that $\phi_{H,C}=\sqrt{2}$ only for dynamical hubs. The last row shows
the boundaries of $A_{H,C}$. \textbf{NA} stands for Not Applicable.}
\label{tab:05}

\begin{tabular}{l c c c c c}
\hline

& \begin{tikzpicture}[auto,node distance = 1.0cm,shorten > = 1pt,> = Stealth,semithick,
 state/.style = {circle, fill=#1, draw=none,text=white},state/.default = gray
                        ]
\node (A) [state]               			{$\pi_i$};
\node (B) [state=white, right of = A]     	{};
\node (C) [state=white,above right of = A]	{};
\node (D) [state=white,below right of = A]	{};
\path[->]   (A) edge ["\hspace{5mm} $P_{m}$"]   (B)
			(A) edge [""]   (C)
			(A) edge [""]   (D);			
\end{tikzpicture}

& \begin{tikzpicture}[auto,node distance = 1.0cm,shorten > = 1pt,> = Stealth,semithick,
 state/.style = {circle, fill=#1, draw=none,text=white},state/.default = gray
                        ]
\node (A) [state]               			{$\pi_i$};
\node (B) [state=white, right of = A]     	{};
\node (C) [state=white,above right of = A]	{};
\node (D) [state=white,below right of = A]	{};
\node (E) [state=white,left of = A]	{};
\path[->]   (A) edge ["\hspace{5mm} $P_{m}$"]   (B)
			(A) edge [""]   (C)
			(A) edge [""]   (D)
			(E) edge [bend left, "$P_{l}$"]   (A)						
			(E) edge [bend right]   (A);						
\end{tikzpicture}

& \begin{tikzpicture}[auto,node distance = 1.0cm,shorten > = 1pt,> = Stealth,semithick,
 state/.style = {circle, fill=#1, draw=none,text=white},state/.default = gray
                        ]
\node (A) [state]               			{$\pi_i$};
\node (B) [state=white, right of = A]     	{};
\node (C) [state=white,above right of = A]	{};
\node (D) [state=white,below right of = A]	{};
\node (E) [state=white,above left of = A]	{};
\node (F) [state=white,below left of = A]	{};
\node (G) [state=white,left of = A]	{};
\path[->]   (A) edge ["\hspace{5mm} $P_{m}$"]   (B)
			(A) edge [""]   (C)
			(A) edge [""]   (D)
			(E) edge [""]   (A)						
			(F) edge []   (A)
			(G) edge ["$P_{l}$ \hspace{5cm}"]   (A);
\end{tikzpicture}

& \begin{tikzpicture}[auto,node distance = 1.0cm,shorten > = 1pt,> = Stealth,semithick,
 state/.style = {circle, fill=#1, draw=none,text=white},state/.default = gray
                        ]
\node (A) [state]               			{$\pi_i$};
\node (B) [state=white, right of = A]     	{};
\node (E) [state=white,above left of = A]	{};
\node (F) [state=white,below left of = A]	{};
\node (G) [state=white,left of = A]	{};
\path[->]   (A) edge [bend left, "$P_{m}$"]   (B)
			(A) edge [bend right]  (B)
			(E) edge [""]   (A)						
			(F) edge []   (A)
			(G) edge ["$P_{l}$ \hspace{5cm}"]   (A);
\end{tikzpicture}

& \begin{tikzpicture}[auto,node distance = 1.1cm,shorten > = 1pt,> = Stealth,semithick,
 state/.style = {circle, fill=#1, draw=none,text=white},state/.default = gray
                        ]
\node (A) [state=white]               			{};
\node (B) [state, right of = A]     	{$\pi_i$};
\node (C) [state=white,above right of = A]	{};
\node (D) [state=white,below right of = A]	{};
\path[->]   (A) edge ["$P_{l}$"]   (B)
			(C) edge [""]   (B)
			(D) edge [""]   (B);			
\end{tikzpicture} 
\\  \hline
{\bf $\mathbf{Type}$}   & \textit{Generator} & \textit{Amplifier} & \textit{Transmitter} & \textit{Attenuator} & \textit{Basin} 
\\  

\multirow{ 2}{*}{{\bf $\mathbf{\phi_{H,C}}$}} & $H_{out}$ & $(H_{out},\sqrt{2})$ & $\sqrt{2}$ & $(H_{in},\sqrt{2})$ & $H_{in}$ \\
& $C_{out}$ & $(C_{out},\sqrt{2})$ & $\sqrt{2}$ & $(C_{in},\sqrt{2})$ & $C_{in}$ \\ 

{\bf $A_{H,C}$}   & \textbf{NA} & $>1$ & $1$ & $<1$ & $0$ 
\\  \hline
\end{tabular}
\end{table}

The last dynamical feature we introduce is the entropy $A_{H}$ and complexity $A_{C}$ amplifications, whose definition is given by Eqs. \ref{equ:a} of Tab. \ref{tab:04}. While $\phi_{H,C}$ give us an idea of which patterns act as pipe-flows, $A_{H,C}$ tell us about the node gain. In other words, the ability to amplifying or diminishing the level of information a pattern
is transmitting. In the case of entropy, for instance, \textit{basin nodes} have zero amplification $A_H=0/H_{in}$. \textit{Transmitter nodes} that equally receive and distribute information would reach up to $A_H=1$ when $H_{out}=H_{in}$, as in the case of \textit{dynamical hubs}. Nodes that increase levels of entropy will have $A_H \gg 1$ and are called {\it amplifier nodes}. In the case of $A_H$, they amplify the level of entropy received from the incoming links taking the system to a disordered one. In the time series, they correspond to short periodic temporal structures followed by irregular patterns. In the case of complexity, an \textit{amplifier node} receive links endowed with either lower or higher entropy, but in both cases outgoing links have higher complexity levels. In the same way, a node might receive higher levels of entropy taking the system down to lower ones. These \textit{attenuator nodes} ($A_H<1$) can be related to changes from irregular fluctuations to more periodic patterns. Likewise, in terms of $A_C$, \textit{attenuator nodes} reduce the complexity levels of the system.

Table. \ref{tab:05} summarizes the node classification based on incoming and departing probabilities $P_l$ and $P_m$. Each dynamical role is characterized by the two fluxes $\phi_{H,C}$ and the two amplifications $A_{H,C}$. 

\section{Results}

\begin{figure}[h]
\centering
\includegraphics[width=\linewidth]{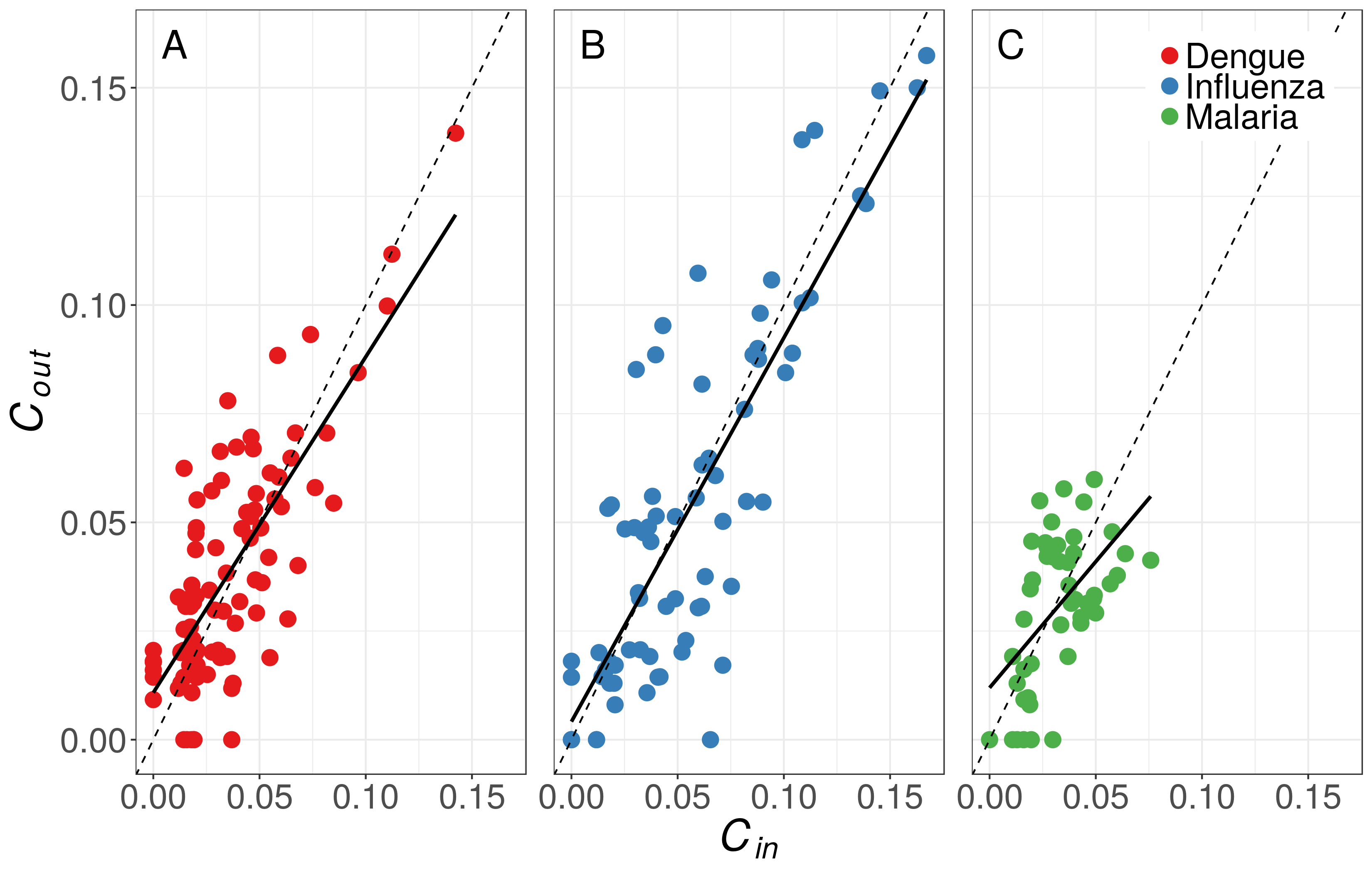}
\caption{Complexity entering a node (pattern) $C_{in}$ vs complexity leaving a node $C_{out}$. Color codes are always the same along the paper: Red for Dengue (A), blue for Influenza (B) and green for Malaria (C). 
Dashed lines correspond to $C_{out}=C_{in}$. 
Solid lines are the regression lines. (A) Dengue coefficient of determination is $R^2=0.6$. (B) Influenza has $R^2=0.71$. (C) In Malaria, $R^2=0.3$.}
\label{fig:03}
\end{figure}

Networks obtained from the same disease are grouped to visualize, first, the incoming and outgoing complexity of a node $i$, respectively,  $C_{in}$ and $C_{out}$ 
(see Eq. \ref{equ:c} of Tab. \ref{tab:03} for details). As it is shown in Fig. \ref{fig:03}, we found positive correlations between these two variables in all 
diseases, which means that the higher the complexity entering a node, the higher the complexity departing from it (note that dashed lines correspond to $C_{in}=C_{out}$). 
Influenza (Fig. \ref{fig:03}B), which is an air-bone disease, is the one reaching higher levels of complexity, with 9 nodes with 
complexities higher than 0.10 (only 2 in the case of Dengue and none in Malaria). 
In contrast, Malaria networks (Fig. \ref{fig:03}C) contain nodes with the lowest amounts of complexity. 

We have obtained the linear regressions accounting for the interplay between $C_{out}$ and $C_{in}$ for the three diseases.
Influenza has the highest coefficient of determination $R^2=0.71$, which means that a linear equation is around 70$\%$ effective to estimate the values of $C_{out}$. 
 Vector-borne diseases have lower $R^2$, in comparison to Influenza. Interestingly, Malaria has the lowest bound with a value of $R^2=0.3$, which reveals the 
 absence of linear correlation between $C_{out}$ and $C_{in}$ . 
 Since dashed lines of Fig. \ref{fig:03} correspond to  $C_{in}$=$C_{out}$, nodes (i.e., patterns)
 lying above these lines are \textit{complexity amplifiers} (see classification of node roles at Tab. \ref{tab:05}), indicating that they distribute more complexity than the levels they receive. 
 On the contrary, nodes below dashed lines 
 act as \textit{complexity attenuators}, since they reduce the complexity existing in previous patterns. 
   
\begin{figure}[htb]
\centering
\includegraphics[width=\linewidth]{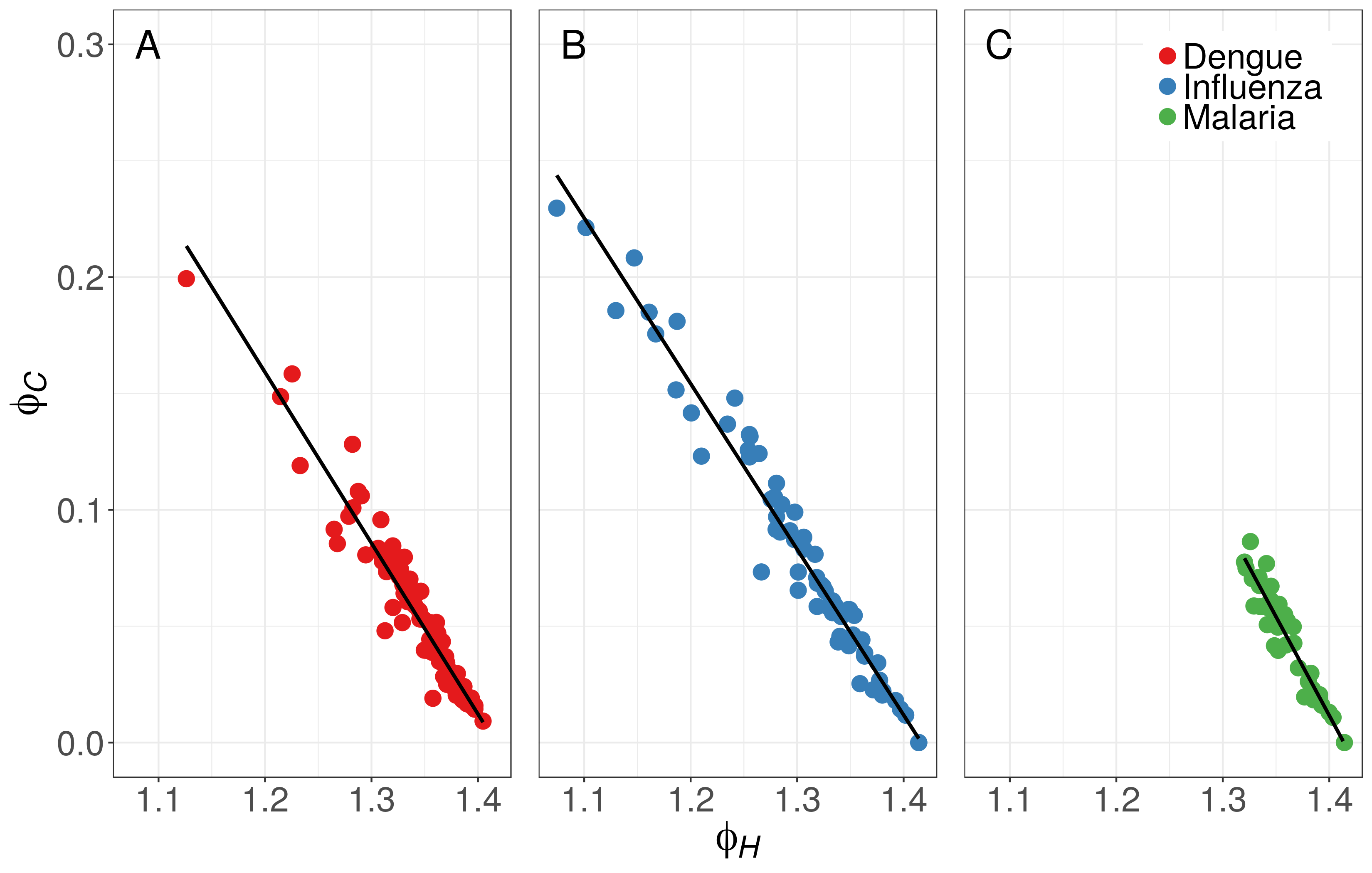}
\caption{Entropy and complexity fluxes and the ($\phi_H,\phi_C$) plane. Solid line correspond to the linear regressions of the three diseases. Coefficient of determinations is $R^2=0.93$ for Dengue (A), $R^2=0.97$
for Influenza (B) and $R^2=0.9$ for Malaria (C).}
\label{fig:04}
\end{figure}

In order to understand how the complexity and entropy of a disease evolve, and what are the patterns that increase or decrease them, we obtain the flux $\phi$ and the amplification $A$ for both
$C$ and $H$ (see Methods for details). Figure \ref{fig:04} shows the interplay between the entropy and complexity fluxes, respectively,  $\phi_H$ and $\phi_C$, that a node handles.
 
Panels of Fig. \ref{fig:04} show the behaviour of the three diseases, revealing negative correlations between entropy and complexity fluxes. This result indicates that we are in a region where the level of stochasticity is high, since, as we can see in Fig. \ref{fig:02}B of the Methods section, entropy and complexity only have positive correlations in situations of high disorder.

Also note that {\it entropy hubs}, i.e., those nodes handling high amounts of entropy, are located in the bottom right side of panels. These nodes receive the highest amount of entropy and distribute it among the rest of patterns in the network, but fail as complexity distributors. Interestingly, Influenza and Malaria have two entropy hubs that reach the highest possible value $(\phi_H=\sqrt{2}$) and, as a consequence, their complexity flux decreases to $\phi_C=0$ (see Methods for an explanation of the boundaries of $\phi_H$ and $\phi_C$).

Similar to Fig. \ref{fig:03}, patterns in Influenza span their 
$\phi_C$ along different levels of $\phi_H$, reaching the highest complexity values. This suggest that some specific patterns in Influenza behave as units of well ``conductance" of complexity, 
while others act as entropy transmitters. On the contrary, Malaria only have entropy hubs, while Dengue seems to be between the other two diseases, having just one complexity hub with a $\phi_C>1.2$. 
Linear regression of the interplay between $\phi_C$ and $\phi_H$ show high values of the coefficient of determination $R^2$ for the three diseases. 

While entropy and complexity fluxes $\phi_{H,C}$ allow to determine the existence of hub patterns and their conductance level, the \textit{amplification} $A$ gives an estimate of how much entropy/complexity a node gains or losses in the network.

\begin{figure}[htb]
\centering
\includegraphics[width=\linewidth]{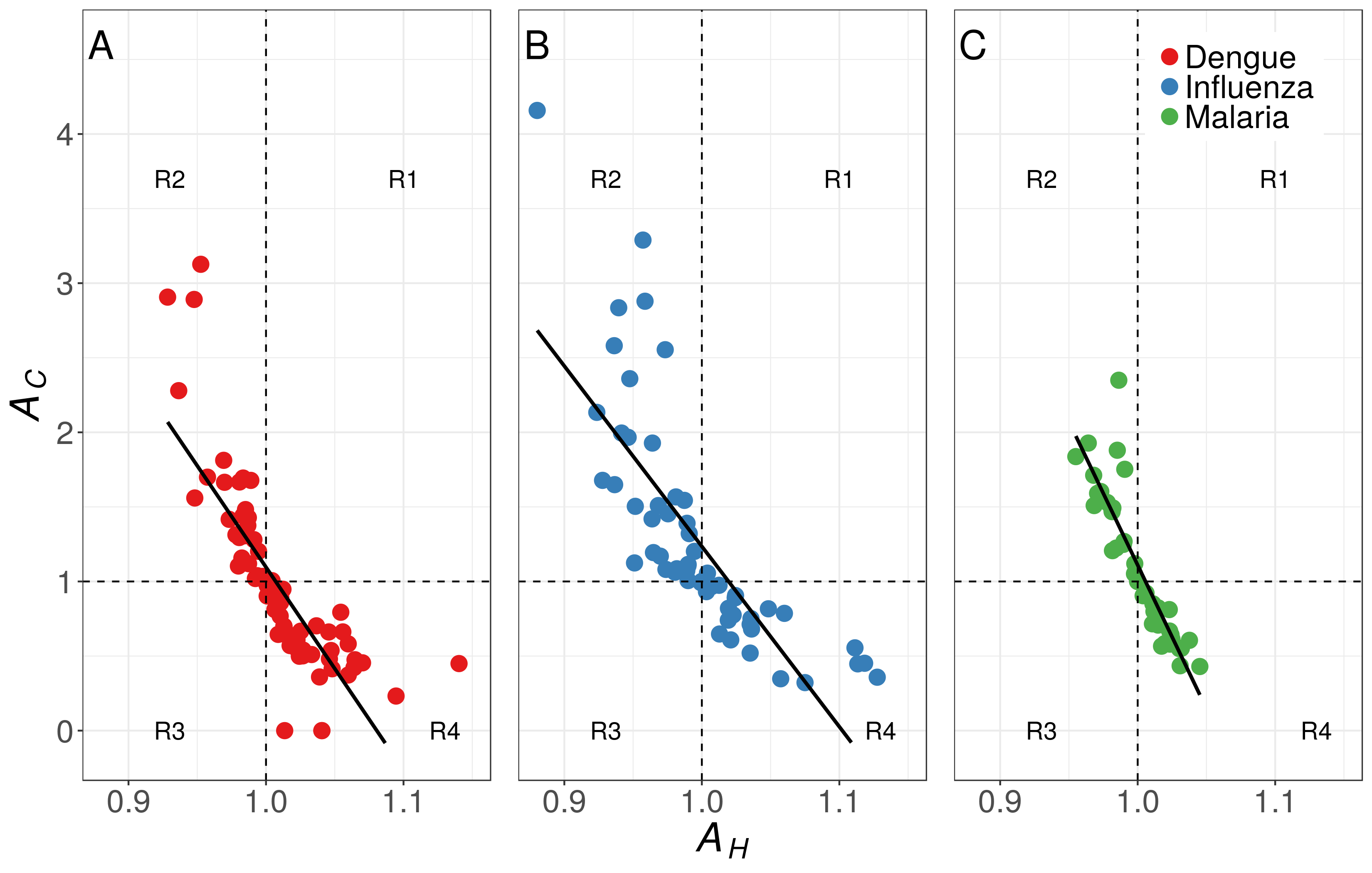}
\caption{Amplifications $A_H \times A_C$ plane. Solid line for model fit. Dashed lines defines region of interest  R1, R2, R3, and R4 allocating nodes of different characteristics. (\textbf{A}) Corresponding coefficient of determinations for Dengue is $R^2=0.66$. (\textbf{B}) Influenza leads to $R^2=0.61$. (\textbf{C}) Malaria: $R^2=0.81$.}
\label{fig:05}
\end{figure}

\textit{Amplification} is defined as the ratio between the incoming and outgoing entropy/complexity that passes through a node $i$ (see Eq. \ref{equ:a} of Table. \ref{tab:04} for the mathematical definition). In Fig \ref{fig:05} we plot ($A_H,A_C$) plane of the three diseases, where we can observe that the amplification parameter allows to define four regions of interest (marked by dashed lines) that, in turn, assign different roles to the nodes of the disease network. Region $R1$ includes nodes that amplify both entropy ($A_H> 1$) and complexity ($A_C> 1$). Since, as we have previously seen, we are in a state where a negative correlation exists between entropy and complexity, there are no nodes lying within this region. Region $R2$ defines nodes that may act as entropy attenuators ($A_H< 1$) and complexity amplifiers ($A_C> 1$). Region $R3$ allocates nodes that attenuate both entropy ($A_H < 1$) and complexity ($A_C< 1$). Finally, region $R4$ correspond to those nodes that increase the entropy ($A_H>1$) while reducing their complexity ($A_C< 1$).

The case of $A_H=A_C=1$ (i.e., no amplification is reported) only occurs for nodes with $\phi_{H,C}=\sqrt{2}$, which is the characteristic of patterns behaving as entropy and complexity hubs.
We can observe in all cases that $C$ faces the most amplification (or attenuation) of its values (see maximum and minimum values of Fig. \ref{fig:05}), while entropy amplification is always bounded between $0.9$ and $1.15$.
In region R2, we observe how Dengue (Fig. \ref{fig:05}A) has three nodes that increase around three times their incoming complexity, while slightly reducing their entropy. However, it is Influenza (Fig. \ref{fig:05}B) that has the node with the higher complexity amplification, which increases the incoming complexity up to 4 times. On the other hand, Malaria (Fig. \ref{fig:05}C) is the disease where the amplification of complexity is the lowest, with only one node with a value higher than 2.

On the other hand, $R4$ allocates nodes that decrease the level of complexity of their incoming patterns and, at the same time, increase their entropy. We can observe that both in Dengue and Influenza, those nodes with the highest entropy amplification $A_H$ depart from the linear behavior that seem to exist in the interplay between $A_C$ and $A_H$.
In fact, when looking at the linear regression, Malaria's coefficient of determination is higher than Dengue and Influenza, but this behavior can be attributed to the deviations from the linear trend that are
reported at both ends of the distributions. In this way, nodes with higher $A_H$ and $A_C$ behave differently from the rest, since in Malaria there are not nodes with extreme values, its coefficient of determination is higher.

\begin{figure}[ht]
\centering
\includegraphics[width=\linewidth,height=13cm]{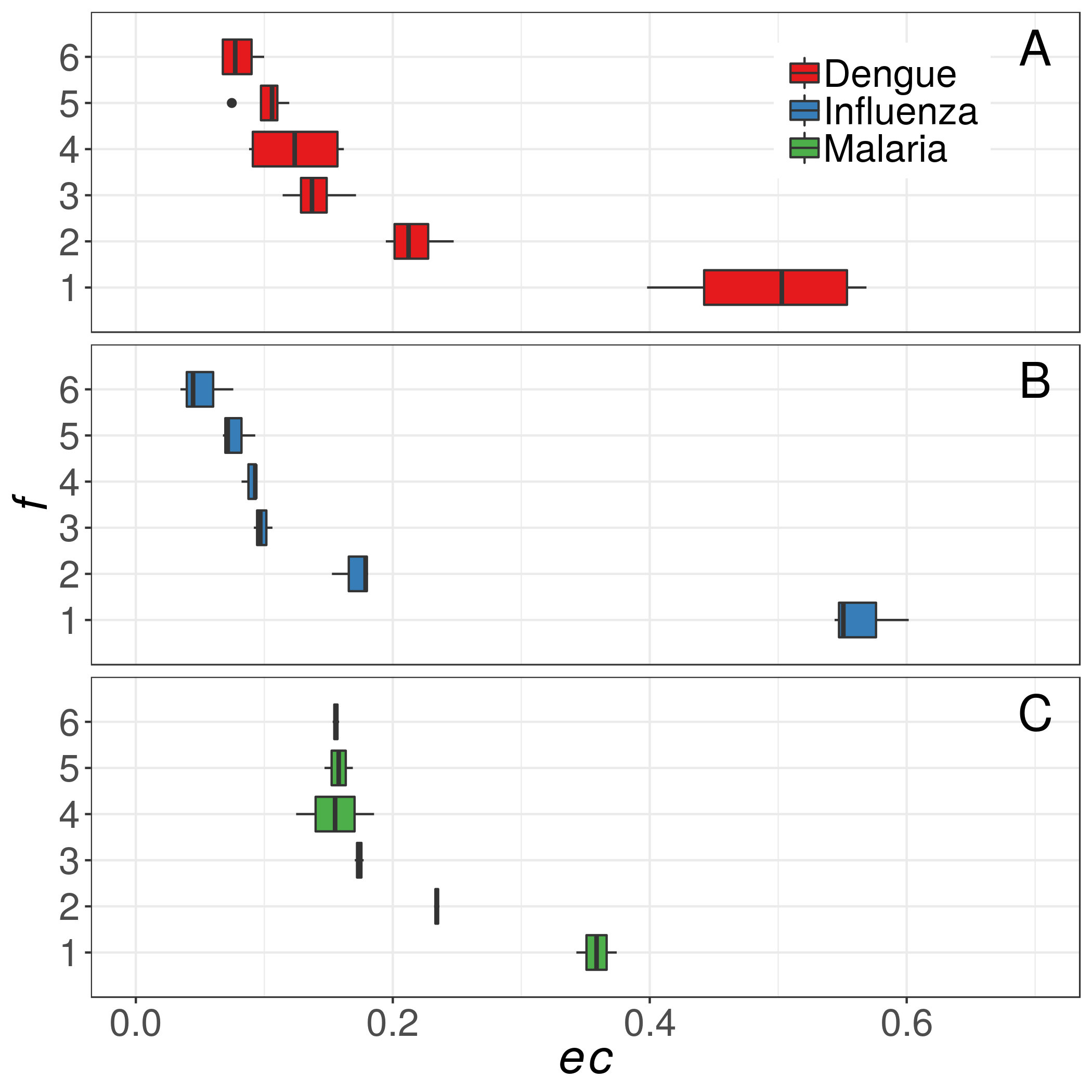}
\caption{Interplay between the entropy/complexity role of the nodes and their centrality in the disease networks.}
\label{fig:06}
\end{figure}

For the sake of a complete characterization, we now pay attention to the interplay between the entropy/complexity role of the nodes and their topological importance in the structure of the networks. With this aim, we use the eigenvector centrality ($ec$) to account for the node importance \cite{newman2010}. $ec$ assumes that central nodes are those that are, at the same time, (i) connected to many nodes and (ii) connected to well connected nodes as well. In parallel with Additionally to $ec$, we obtain the \textit{pattern fluctuation} ($f$), which quantifies the variability inside each pattern: those patterns whose elements increase and decrease one after the other, will have a high $f$, while those patterns that monotonically increase or decrease, will have the lowest $f$ (see Methods). For each disease, assuming a pattern dimension of $D=4$, we obtain $24$ different patterns, whose variability $T$ results in 6 different levels of fluctuation. The higher the variability $T$, the higher the internal disorder of a pattern $\pi$ and the higher its $f$. Table \ref{tab:06} of the Supp. Info. shows how patterns with $D=4$ are grouped in $6$ different values of $f$. Figure \ref{fig:06} depicts the fluctuation $f$ vs the topological importance $ec$, where nodes grouped in terms of its $f$. We can observe that central (important) nodes are those with low levels of fluctuation. In other words, patterns with low variability are the central ones in the network structure. On the contrary, as the internal fluctuation of a pattern increases, its probability of being a hub decreases. As a consequence, peripheral nodes are associated with patterns having the highest fluctuations.

It is worth noting that for Dengue and Influenza,  nodes with the lowest fluctuations, i.e., those patterns associated to either a monotonic increase or decrease of the number of infected individuals, are those with the highest centrality. This fact is probability caused by the existence of abundant periods where the number of infected individuals monotonically increase or decrease, leading to a high number of appearances of patterns
$(0,1,2,3)$ or $(3,2,1,0)$, which increases their number of links and, unavoidably, their eigenvector centrality. At the same time, when a node accumulates great part of the centrality, it relegates the rest of the nodes
of the network to a secondary role. However, Malaria behaves in a different way, since we can observe how the heterogeneity of the eigenvector centrality is not that high as in the other two diseases.

Next, we compare the previous results about the interplay between the node role and its centrality with those obtained with synthetic time series. The reason is to qualitatively observe whether dynamical signatures of these diseases are similar to those of classical models. With this aim, we simulated six models of different levels of complexity and entropy. Specifically:
\begin{itemize}
\item A Linear Gaussian Process (LGP, $\mathcal{N}(0,1)$), with the aim of having a signal with the highest entropy, and, as a consequence, with the lowest complexity. 
\item A second-order auto-regressive model (AR(2), $x_{t+2}=0.7x_{t+1}+0.2x_t+\epsilon_t$) with $\epsilon_t$ being a Gaussian noise. 
\item A Self-Exciting Threshold AR model (SETAR($k;p_1,p_2$)), with $k$ as the number of regimes, and $p_1, p_2$ the order of the autoregressive parts.
SETAR models have been largely used when modeling ecological systems that oscillate between two non-linear regimes with different delays. This model accounts for higher levels of complexity by decreasing its entropy levels. We modeled it by the SETAR(2;2,2) switching between regimes ($0.62+1.25x_{t-1}-0.43x_{t-2}+0.0381\epsilon_t$) if $x_{t-2}\leq3.25$ and ($2.25+1.52x_{t-1}-1.24x_{t-2}+0.0626\epsilon_t$), otherwise. 
\end{itemize}

We also modeled three chaotic systems: 
\begin{itemize}
\item A logistic map: $x_{t+1}=4x_t(1-x_t)$.
\item A R\"osler  system: $\dot{x}=-y-z$, \hspace{0.2cm} $\dot{y}=x+0.2y$, \hspace{0.2cm} $\dot{z}=0.2+z(x-5.7)$.
\item A  Lorenz system: $\dot{x}=10(y-x)$, \hspace{0.2cm} $\dot{x}=x(28-z)-y$, \hspace{0.2cm} $\dot{x}=xy-2.6667z$.
\end{itemize}

\begin{figure}[ht]
\centering
\includegraphics[width=\linewidth]{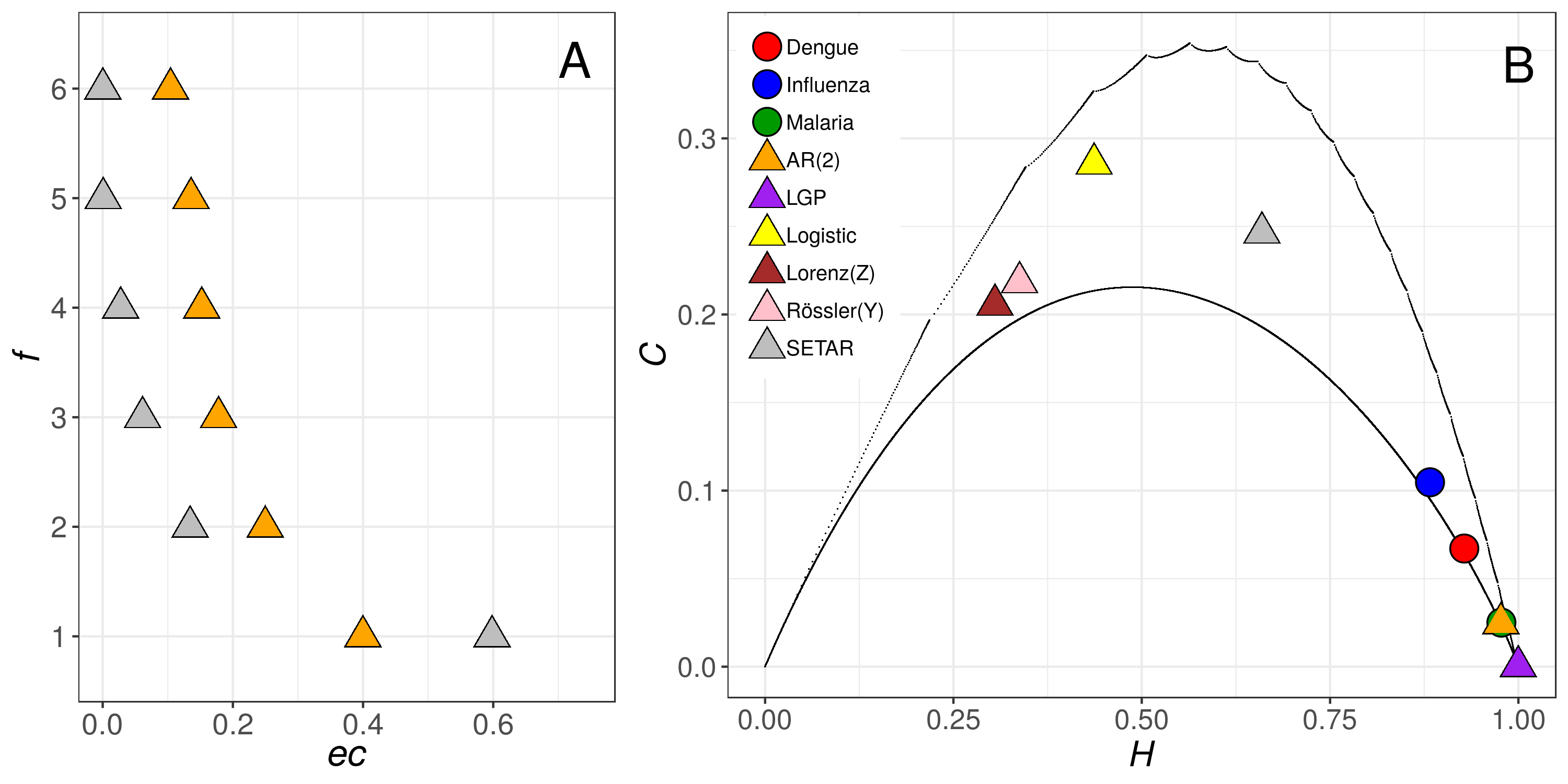}
\caption{Comparison between real datasets and synthetic models 
Colors correspond to AR(2) (orange), Logistic (purple), Lorenz (variable z) (yellow), LGP (brown), Rösler (variable y) (pink) and  SETAR (gray) models. (A) SETAR and AR(2) share similitudes with the behaviour reported in  seasonal and vector-borne diseases (see Figure \ref{fig:06}). (B) ($H,C$) plane for all synthetic models and real datasets. Black lines represent theoretical curves of maximum and minimum complexity for $D=4$.}
\label{fig:07}
\end{figure}

We set the length of the time series generated by each model to $M=10^4$ after removing the first 1000 samples to avoid possible transients. We reconstructed the respective symbolic networks with $D=4$ and calculated both their patterns' centralities and fluctuations. Comparing the dynamics vs topology diagrams, we detected that AR(2) and SETAR models behave similarly to the diseases showed in Fig. \ref{fig:06}. We observed how the SETAR model has a considerable gap between peripheral and central ones (Fig. \ref{fig:07}A), something that was already observed for the Influenza networks (Fig. \ref{fig:06}B). At the same time, the AR(2) model shows a smoother decay combined with higher values of centrality for all patterns (Fig. \ref{fig:07}A), as it is the case of Malaria (Fig. \ref{fig:06}C).
In both cases, peripheral nodes are the ones with the highest fluctuations, while hubs correspond to patterns having internal monotonic increase/decrease.

We now reproduce the ($H,C$) plane of global measures for all synthetic models and compare them with the experimental results (see Fig. \ref{fig:02}B of Materials and Methods). 
Figure \ref{fig:07}B depicts the ($H,C$) plane accounting for the interplay between these two variables and giving interesting information about the global organization of each disease dynamics. Here, the properties of the time series are captured with the distribution of all patterns appearing in the time series for $D=4$ and visualized in the ($H,C$) phase diagram. All models are enclosed in-between the theoretical maximum and minimum complexity described by black curves. These curves depend solely on the functional form that describes entropy and disequilibrium, as well as the dimension of the space of probabilities associated to the system. Specifically, acquiring the extreme values (maximum or minimum) of complexity is an optimization problem where we compute the set of probability distributions that optimizes the disequilibrium due to an specific value of entropy. The previous makes these curves useful when comparing the dynamics of time series of different nature and/or even lengths, in the same ($H,C$) space but with the same dimension.

The R\"osler (variable y), Lorenz (variable z) and Logistic chaotic maps appear in the region of positive correlations between $H$ and $C$, which corresponds to high levels of complexity, which are, in turn, correlated with the low levels of entropy. Note that the region before the maximum showed by the theoretical curves, can be considered 
as the division between ordered and disordered states and crucially determines the kind of correlation between entropy and complexity.
In the case of autoregressive models, all of them lie in the region of negative correlations. SETAR depicts a high level of complexity with a tendency to the disorder, something that is expected for highly non-linear coupled systems with Gaussian noise. As soon as noise begins to drive the dynamics, and the nonlinearity vanishes, which is the case of AR(2), complexity decreases while entropy increases. In the extreme case of the LGP, its disorder is the highest and its complexity goes to zero.

Finally, it is important to highlight that fluctuations are valid to achieve the types of $H\times C$ diagrams, as well as for identifying types of nodes. In the Supplementary Information, we reproduced the results of the relationships between ($C_{in}$, vs $C_{out}$), ($\phi_{H}$, vs $\phi_{C}$) and ($A_{H}$, vs $A_{C}$). Without loss of generality, figures of supplementary material are consistent with the behaviour observed in Figs. \ref{fig:03}, \ref{fig:04}, \ref{fig:05}.

\section{Discussion}
In this work we investigated the problem of how entropy and complexity flow along the temporal evolution of different epidemic diseases. Time series containing the number of individuals infected with Dengue, Influenza and Malaria are encoded by ordinal patterns, i.e. symbols that are connected to each other sequentially in order to build symbolic graphs. To measure how entropy and complexity flow along ordinal patterns, we endowed symbolic networks with patterns (nodes) that have the ability of transferring their amounts of entropy/complexity to other patterns. Consequently, nodes, or ordinal patterns, might amplify or attenuate the entropy/complexity transmitted to other nodes.

In this way we propose a family of five parameters to quantify how the entropy/complexity flows among temporal patterns. Two of them quantify how the information arrives (departs) to (from) a pattern in terms of its entropy and complexity. They are normalized allowing quantitative comparisons between different diseases and synthetic models. Other two features quantify both the level of conductance (flux), and the gain of information of a pattern (amplification). The last one (fluctuation), assesses the inner variability of the dynamics inside each pattern. 

Using these metrics, we characterize disease's outbreaks converting epidemiological cohorts into symbolic networks, where temporal fluctuations of signals might communicate among each other. 
The aim of this approach is to unveil whether the exchange of information among each signal's patterns shows differences among different diseases.

One may consider the incoming/outgoing complexity as information that enters/leaves a given pattern that, in turn, consists on a dynamical state of the disease prevalence given by the combination of $D$ sequential values, accounting the number of infected individuals. In this context, we have seen in Fig. \ref{fig:03} how the incoming and outgoing complexity are positively correlated for the 
three diseases. In this way,  the more information a signal's segment receives from a previous one, the more information it sends to following temporal structures. From the three diseases, Influenza is the one reaching the highest correlation. In contrast, Malaria is the disease with the lowest correlation, suggesting the influence of another variables in the transmission of complexity levels.

To the best of our knowledge, this works present the first evidence of the link between the inter dynamics of ordinal patterns and the role it plays in a symbolic network. Thus, fluctuation ($f$)  seems an interesting quantity when measuring the internal dynamics of temporal structures in a signal. By associating node fluctuation with its structural importance, we discover how hub patterns correspond to those nodes with the lowest fluctuation (i.e., hubs are nodes whose related patterns behave more monotonically). Reciprocally, the more peripheral a node is, the more fluctuations the patterns has. Interestingly, this behaviour is reported in all diseases. However, the distribution of centrality in Malaria is more homogeneous, with the centrality of both hubs and peripheral nodes
between $0.1$ and $0.4$ (Fig. \ref{fig:03}C). On the other side, the distribution of centrality in Dengue and, specially, Influenza, is quite heterogeneous, with few nodes acquiring a centrality above $0.5$ (Fig. \ref{fig:03}A-B). This ``accumulation'' of centrality might be driven by the seasonal nature of Influenza, and seasonal-like nature of the Dengue data for some of the countries considered. Nodes with low $f$ appear regularly in the time series.
Importantly, when nodes are grouped in terms of their variabilities (see Sup. Info.), fluctuation shows comparable results to those that were disclosed by flux, amplitude and nodal complexity.
For synthetic time series, we detected qualitative similarities between Influenza and nonlinear autoregressive models, while observing that Malaria resembles more closely to linear AR processes; this relations were confirmed by a simple analysis of the entropy-complexity plane.

It is worthwhile to highlight that the family of novel nodal features is not only valid for symbolic networks, but for any kind of weighted-directed graphs. 
When having directed and weighted networks, e.g., in metabolic, genetic or systems biology scenarios, one can translate it to directed graphs by repeating the methodology proposed in this paper. 

Nonetheless, further research is needed to better understand both the structure and dynamics of symbolic networks. On one hand, the 
construction of symbolic networks from time series should be evaluated when patterns contain delays $\tau\neq 1$. Additionally, working 
with large time series is desirable and would allow arriving to larger
 dimensions $D$ for the patterns length. Hence, obtaining large size connected networks would benefit its statistical properties. 
 Apart from that, a detailed analysis of the robustness of this kind of networks is still remaining and
 would help understanding the global properties of these particular graphs.

On the other hand, one of the weakness of our study was the poor quality of the available data. Unfortunately, the lack of organization in the acquisition and 
storage of surveillance data from developing countries makes difficult to obtain deeper conclusions for healthy public policies. 
Datasets usually lack of complete information. For example, there exists few weeks with zero reported cases in countries where it is well known the continuous 
presence of infected population. In addition, epidemiological weeks do not reflect the real case of infected individuals. Due to human migrations, infected 
subjects in a country could have been infected in foreign ones (imported cases), which is a problem when performing studies of spatio-temporal 
differentiation. At the moment of retrieving the data, there were no other countries with available datasets of the three diseases we consider, and surprisingly, 
the Influenza is not well documented in most cases. In addition, although Dengue and Malaria belong to vector-borne diseases, both of
 them have different strains of virus. Hence, a study with high-quality datasets would find differences among them, but it's mandatory 
 to have large prevalence records of both strains to large enough signals. In general, works with datasets of different nature, would 
 shed light in how different/similar real systems are by means of these entropy/complexity features. 

All in all, this work opens the door to new experimental designs to extract information from time series, as well as from directed 
and weighted networks. However, this methodology should be used as a complementary tool analyzing time series of real datasets. In general, the use of 
these metrics and its related methodologies will grant new information to better understand 
the interplay between temporal structures in natural and artificial collection of samples of finite nature.


\dataccess{Diseases data available at: \href{https://github.com/JohannHM/Disease-Outbreaks-Data.git}{GitHub repository. Disease-Outbreaks-Data}.}
\aucontribute{JLHD and JHM worked on acquisition of data, JMB and JHM on the conception and experimental design, JLHD, JHM, MC and JMB worked on analysis and interpretation of data, JHM drafting the article, All authors equally collaborated on writing the final manuscript and revising it critically for intellectual content.}
\competing{We declare we  have no competing interests.}
\funding{JHM thank to FAPESP grant 2016/01343-7 for funding my visit to ICTP-SAIFR from March-April 2019 where part of this work was done. JLHD was supported by the S\~ao Paulo Research Foundation (FAPESP) under Grants No. 2016/01343-7 and No. 2017/05770-0. JMB is founded by MINECO (project FIS2017-84151-P)}
\ack{JHM thanks to Xavier Bosch-Capblanch, Swiss Tropical and Public Health Institute. University of Basel, and C. Mart\'inez for valuable conversations}

\vskip2pc

\bibliographystyle{RS} 
\bibliography{epidemias_PRSA_v8}  


\appendix
\section*{Supplementary information}
\subsection*{Fluctuations for $D=4$}

In Tab. \ref{tab:02} of Methods section, we introduce an easy-to-take list of fluctuations for the trivial case when patterns have $D=3$. However, for the results shown in this work, we take consider $D=4$, which leads to networks of 24 possible different patterns. From them, we group patterns $\pi$ with the same variability $T$ and assign them the same fluctuation $f$.

Table \ref{tab:06} shows the 24 patterns distributed into the 6 different fluctuations. Patterns are sorted in increasing order of fluctuation. The higher the variability $T$, the higher its fluctuation. Another way of thinking about fluctuations is based on its disorder level: The higher the fluctuation is, the closer the pattern is to disorder.

\begin{table}[ht]
\centering

\begin{tabular}{cccc}

\begin{tabular}{|l|c|c|}
 \hline
 {\bf $\mathbf{\pi}$}&  {\bf $\mathbf{T}$} & {\bf $\mathbf{f}$} \\  \hline
 $0,1,2,3$& 4.243 & 1 \\ 
 $3,2,1,0$& 4.243 & 1 \\ 
 $0,1,3,2$& 5.064 & 2 \\ 
 $1,0,2,3$& 5.064 & 2 \\ 
 $2,3,1,0$& 5.064 & 2 \\ 
 $3,2,0,1$& 5.064 & 2 \\ \hline
\end{tabular}

\hspace{0.cm}

\begin{tabular}{|l|c|c|}
 \hline
 {\bf $\mathbf{\pi}$}&  {\bf $\mathbf{T}$} & {\bf $\mathbf{f}$} \\  \hline 
 $0,2,1,3$& 5.886 & 3 \\ 
 $0,2,3,1$& 5.886 & 3 \\ 
 $1,3,2,0$& 5.886 & 3 \\ 
 $2,0,1,3$& 5.886 & 3 \\ 
 $3,1,0,2$& 5.886 & 3 \\ 
 $3,1,2,0$& 5.886 & 3 \\ \hline
\end{tabular}

\hspace{0.cm}

\begin{tabular}{|l|c|c|}
 \hline
 {\bf $\mathbf{\pi}$}&  {\bf $\mathbf{T}$} & {\bf $\mathbf{f}$} \\  \hline
 $0,3,2,1$& 5.991 & 4 \\ 
 $1,0,3,2$& 5.991 & 4 \\ 
 $1,2,3,0$& 5.991 & 4 \\ 
 $2,1,0,3$& 5.991 & 4 \\ 
 $2,3,0,1$& 5.991 & 4 \\ 
 $3,0,1,2$& 5.991 & 4 \\ \hline 
\end{tabular}

\hspace{0.cm}

\begin{tabular}{|l|c|c|}
 \hline
 {\bf $\mathbf{\pi}$}&  {\bf $\mathbf{T}$} & {\bf $\mathbf{f}$} \\  \hline
 $0,3,1,2$& 6.812 & 5 \\ 
 $1,2,0,3$& 6.812 & 5 \\ 
 $2,1,3,0$& 6.812 & 5 \\ 
 $3,0,2,1$& 6.812 & 5 \\ 
 $1,3,0,2$& 7.634 & 6 \\ 
 $2,0,3,1$& 7.634 & 6 \\ \hline
\end{tabular}

\end{tabular}
\caption{Patterns, variability $T$ and the corresponding fluctuation $f$ for $D=4$. First column: $\pi$ patterns. Second column: Internal variability $T$. Third column: The corresponding fluctuations $f$.}
\label{tab:06}
\end{table}

\subsection*{Retrieving node information with fluctuations}

In the following figures we reproduce the main results of the paper in terms of fluctuations. We can observe how fluctuation captures the same behaviour already obtained en Figs. \ref{fig:03}, \ref{fig:04} and \ref{fig:05}.

\begin{figure}[ht]
\centering
\includegraphics[width=0.8\linewidth]{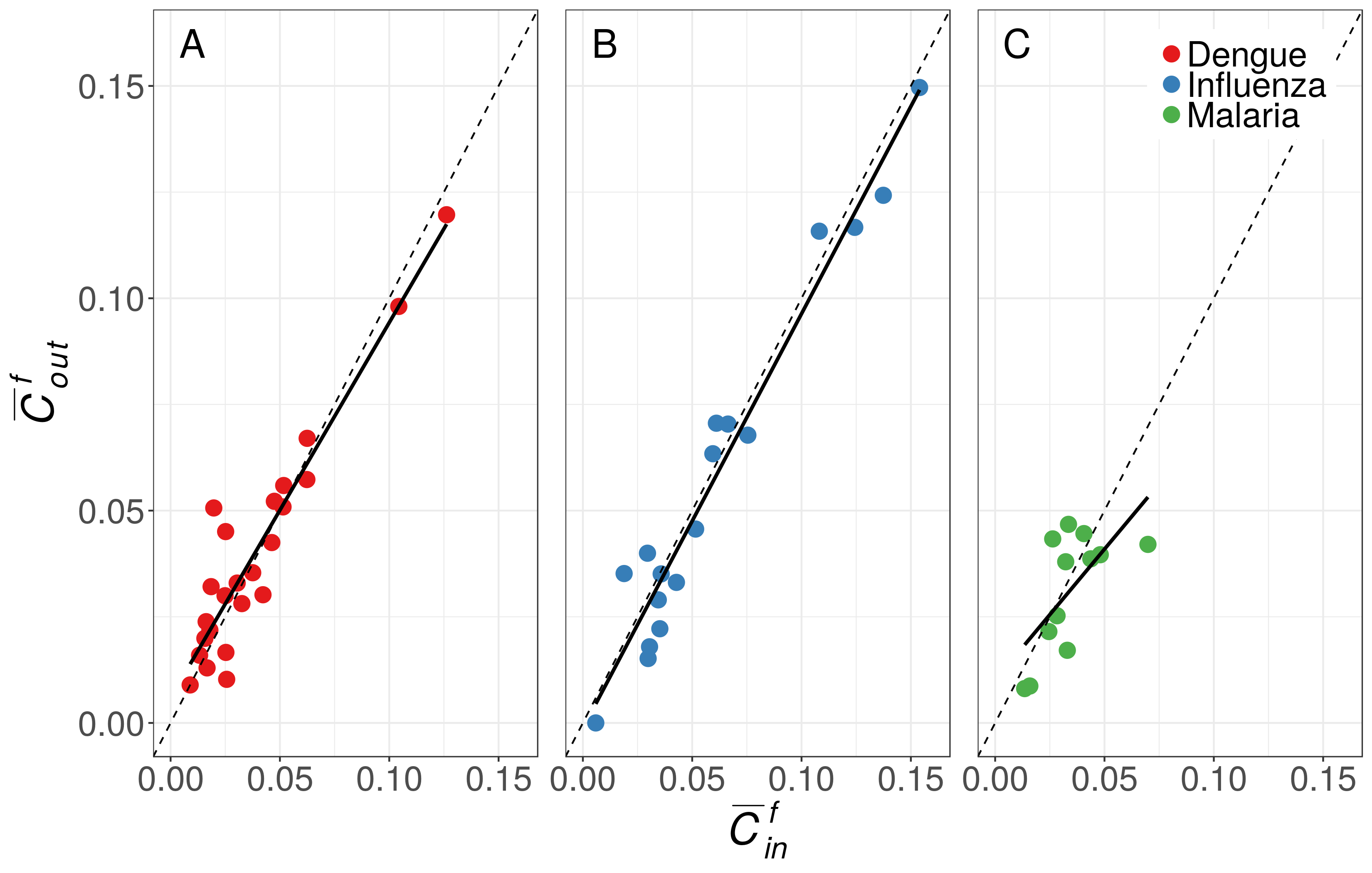}
\caption{Average complexity of nodes grouped by fluctuation $\bar{C}_{in,out}^f$. Solid line correspond to the linear fits. Dashed line is $\bar{C}_{out}^f=\bar{C}_{in}^f$. Coefficient of determination are: $R^2=0.88$ for Dengue (A),  $R^2=0.95$ for Influenza (B) and $R^2=0.44$ for Malaria (C).}
\label{fig:si03}
\end{figure}

In Figure \ref{fig:si03} we computed $C_{in,out}$ for all nodes. We grouped these values in terms of fluctuations. For patterns with the same fluctuation, we obtain its mean value $\bar{C}_{in,out}^f$ and we plot both variables. Similar to Fig. \ref{fig:03}, Influenza is the disease that reaches the highest level of complexity.

\begin{figure}[ht]
\centering
\includegraphics[width=0.8\linewidth]{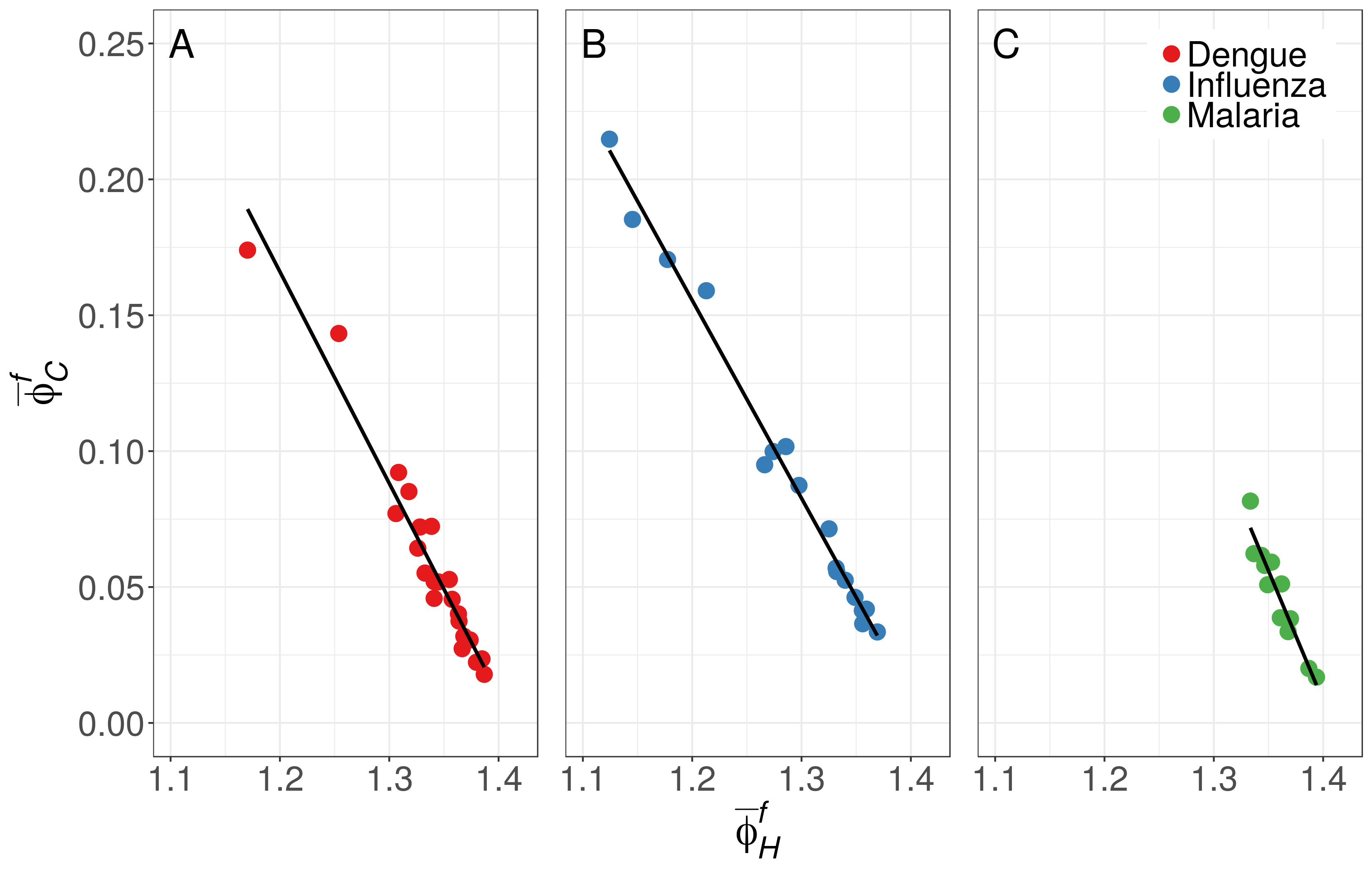}
\caption{Average flux $\bar{\phi}_{H,C}^f$ of nodes grouped by fluctuation $f$. Solid lines correspond to the linear fits. Coefficient of determination are: $R^2=0.95$ for Dengue (A),  $R^2=0.99$ for Influenza (B) and 
$R^2=0.91$ for Malaria (C).}
\label{fig:si04}
\end{figure}

Figure \ref{fig:si04} corresponds to Fig. \ref{fig:04} of the main text. 

\begin{figure}[ht]
\centering
\includegraphics[width=0.8\linewidth]{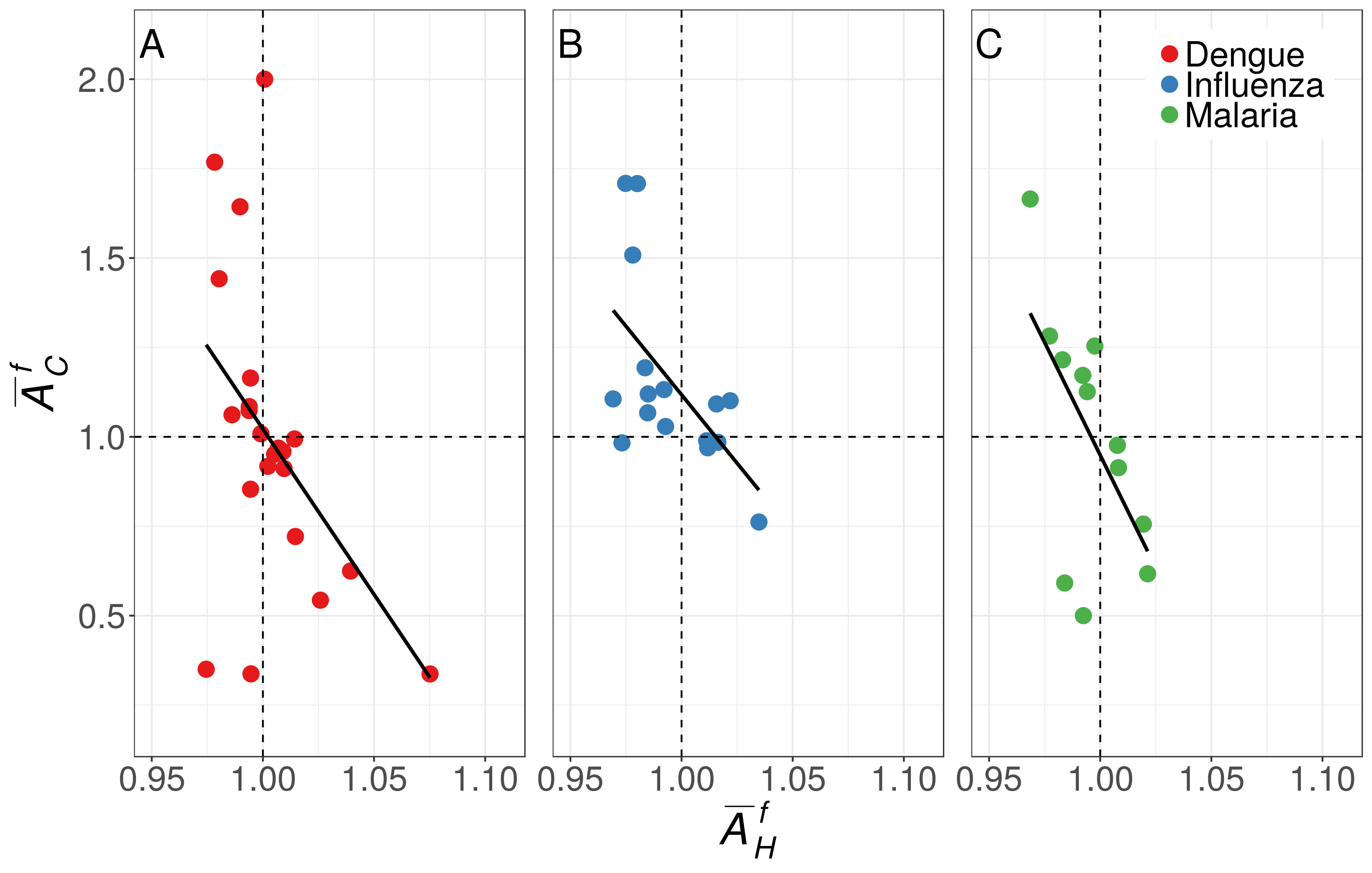}
\caption{Average amplification $\bar{A}_{H,C}^f$ of nodes grouped by fluctuation $f$.  Solid line correspond to the linear regressions. Coefficient of determinations are: $R^2=0.22$ for Dengue (A),  
$R^2=0.34$ for Influenza (B) and $R^2=0.36$ for Malaria (C).}
\label{fig:si05}
\end{figure}

Figure \ref{fig:si05} corresponds to Fig. \ref{fig:06} of the main text. The negative correlation between $A_C$ and $A_H$ is also reported.

\end{document}